\documentclass[%
 reprint,
superscriptaddress,
nofootinbib,
 amsmath,amssymb,
 aps,
pre,
floatfix
]{revtex4-1}

\usepackage{graphicx}
\usepackage{xspace}
\usepackage[normalem]{ulem}
\graphicspath{{figures/}} 

\usepackage{dcolumn}
\usepackage{bm}
\usepackage{comment} 
\usepackage[colorlinks=true,allcolors=blue]{hyperref}

\usepackage{physics}
\usepackage{color}

\newcommand \brandeis {Martin Fisher School of Physics, Brandeis University,\\
415 South Street, Waltham, Massachusetts 02453, USA
}

\newcommand \tufts {Department of Physics and Astronomy, Tufts University, \\ 574 Boston Avenue, Medford, Massachusetts 02155, USA}

\newcommand \rochester {School of Physics and Astronomy, Rochester Institute of Technology, Rochester, NY 14623}

\newcommand*{\Ri}{R_\text{i}}
\newcommand*{\Riw}{\Ri/W}

\def\DE{D_\text{E}}
\def\Dr{D_\text{r}}

\newcommand*{\EA}{E_\text{A}}
\newcommand{\ie}{\textit{i.e.}\xspace}

\newcommand{\etal}{et al.\xspace}

\newcommand*{\tf}{t_\text{f}}

\def\phd{+\frac{1}{2}}
\def\mhd{-\frac{1}{2}}
\newcommand*{\nyexp}{\expval{|n_y|}_{y,t}}
\newcommand*{\xp}{x_\text{p}}
\newcommand*{\xdecay}{x_\text{d}}
\newcommand{\Deltany}{\Delta n_y}
\begin{document}


\title{From Disks to Channels: Dynamics of Active Nematics Confined to an Annulus}
\author{Chaitanya Joshi}%
\email{chaitanya.joshi@tufts.edu}
\affiliation{\tufts}
\affiliation{\brandeis}%
\author{Zahra Zarei}

\affiliation{\brandeis}
\author{Michael M. Norton}
\affiliation{\rochester}%


\author{Seth Fraden}
\affiliation{\brandeis}%
\author{Aparna Baskaran}
\email{aparna@brandeis.edu}
\affiliation{\brandeis}%
\author{Michael F. Hagan}
\email{hagan@brandeis.edu}
\affiliation{\brandeis}%


\date{\today}

\begin{abstract}
Confinement can be used to systematically tame turbulent dynamics occurring in active fluids. Although periodic channels are the simplest geometries to study confinement numerically, the corresponding experimental realizations require closed racetracks. Here, we computationally study 2D active nematics confined to such a geometry --- an annulus. By systematically varying the annulus inner radius and channel width, we bridge the behaviors observed in the previously studied asymptotic limits of the annulus geometry: a disk and an infinite channel. We identify new steady-state behaviors, which reveal the influence of boundary curvature  and its interplay with confinement. We also show that, below a threshold inner radius, the dynamics are insensitive to topological constraints imposed by boundary conditions. We explain this insensitivity through a simple scaling analysis. Our work sheds further light on design principles for using confinement to control the dynamics of active nematics.

\end{abstract}

\pacs{Valid PACS appear here}
\maketitle


\section{\label{sec:level1}Introduction}

 Active matter describes a collection of constituents that consume energy at the level of the individual units to generate motion. Interactions between the units combined with their motion result in emergent behaviors that span scales much larger than those of the particles or their interactions, and would be thermodynamically forbidden in equilibrium materials \cite{Marchetti2013,Ramaswamy2010,Toner2005}.
Suspensions of such active units, termed active fluids, are a paradigm to describe different biological systems such as the cytoskeleton of cells \cite{Julicher2007},  bacterial colonies \cite{Wensink2012} and tissues and cell sheets  \cite{Duclos2017,Saw2017,Kawaguchi2017}. Enabled by the continuous conversion of chemical energy into mechanical work due to biomolecular processes, these systems exhibit myriad emergent functionalities that are crucial for their organisms' lifecycles, such as motility, division, self-healing, and morphogenesis \cite{Giomi2014,Maroudas-Sacks2020,Vafa2022,Khoromskaia2023,Guillamat2022,Hoffmann2022}. 

Active nematic liquid crystals, which are composed of motile energy-consuming anisotropic units, provide a promising platform to achieve similar capabilities in synthetic or biomimetic systems \cite{Doostmohammadi2018}. However, bulk active nematics exhibit chaotic turbulent-like flows that lack long-range order \cite{Dombrowski2004,Cisneros2007,Wolgemuth2008,Creppy2015,Giomi2015,Blanch-Mercader2018,Peng2021}. 
Thus, they are unable to perform functions such as generating work or driving net material transport without a means to suppress this turbulence.

Confinement can control active flows \cite{Voituriez2005,Duclos2018,Theillard2017,Woodhouse2012,Wu2017,Varghese2020,Chandrakar2020,Edwards2009, Giomi2011, Giomi2012, Wioland2013,Ravnik2013, Doostmohammadi2016, Wioland2016, Shendruk2017,Gao2017, Doostmohammadi2017,VanTeeffelen2009}, and enable harnessing them for transport and other functions (e.g. \cite{Furthauer2012,Thampi2016,Li2021}). The simplest confinement in 2D is that of an infinite channel of finite width, computationally implemented using periodic boundary conditions. It has been shown that, depending on the channel width and activity level, active nematics in such channels exhibit a variety of emergent states including `dancing' defect pairs and coherent flow along the channel \cite{Shendruk2017,Hardouin2019}. However, in experiments, such a channel can only be mimicked by physically joining its two ends, leading to a curved racetrack or annulus geometry \cite{Wu2017,Opathalage2019,Hardouin2022,Hardouin2020a}, or by finite-length `lanes' in which end-effects may arise \cite{Hardouin2019,Opathalage2019}.
\begin{figure}[t]
\centering
\includegraphics[width=0.9\linewidth]{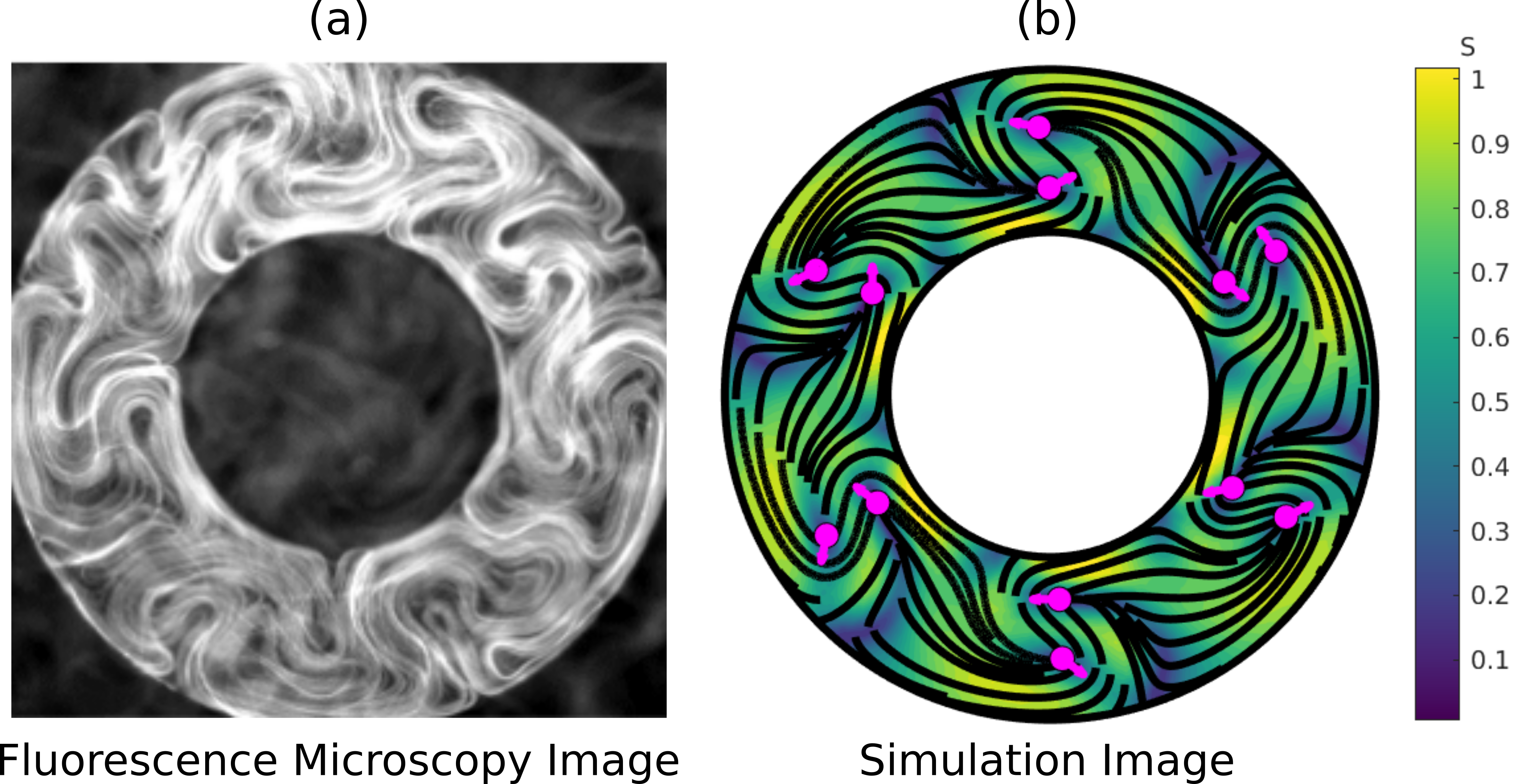}
\caption{\textbf{ 2D active nematics confined to an annulus geometry} (a) Snapshot of the microtubule-kinesin active nematics system, obtained using fluorescence microscopy. (b) Snapshot from a hydrodynamic simulation of the same system. The lines indicate the director field lines while the color corresponds to the scalar order parameter $S$. 
The magenta arrows indicate the positions and orientations of $\phd$ defects.}
\label{fig:annulus}
\end{figure}
For instance, Fig. \ref{fig:annulus} shows a widely studied active matter system of cytoskeleton-based 2D active nematics \cite{Sanchez2012,Doostmohammadi2018} confined to an annulus geometry. In bulk, this system exhibits the widely studied phenomenon of low Reynolds number turbulence \cite{Giomi2015,Thampi2016,Thampi2016a, Guillamat2017, Lemma2019,Doostmohammadi2017,Opathalage2019,Alert2020,Alert2022,Martinez-Prat2019,Martinez-Prat2021}. Experiments in the annulus geometry \cite{Hardouin2022,Hardouin2020a} and lane configurations \cite{Hardouin2019} demonstrate that confinement can tame this turbulence to generate ordered flows. However, despite these experiments and previous theoretical investigations of active nematics confined in channels \cite{Shendruk2017,Duclos2018,Hardouin2019,Varghese2020,Chandragiri2019,Chandragiri2020,Li2021,Wagner2022}, disks \cite{Norton2018,Opathalage2019,Hardouin2022,Mirantsev2021}, and annulus geometries \cite{Shendruk2017,Chen2018,Hardouin2022,Hardouin2020a,deOliveira2023}, the influence of channel curvature and finite channel length on emergent behaviors has yet to be studied in detail. 

\begin{figure}[t]
\centering
\includegraphics[width=0.9\linewidth]{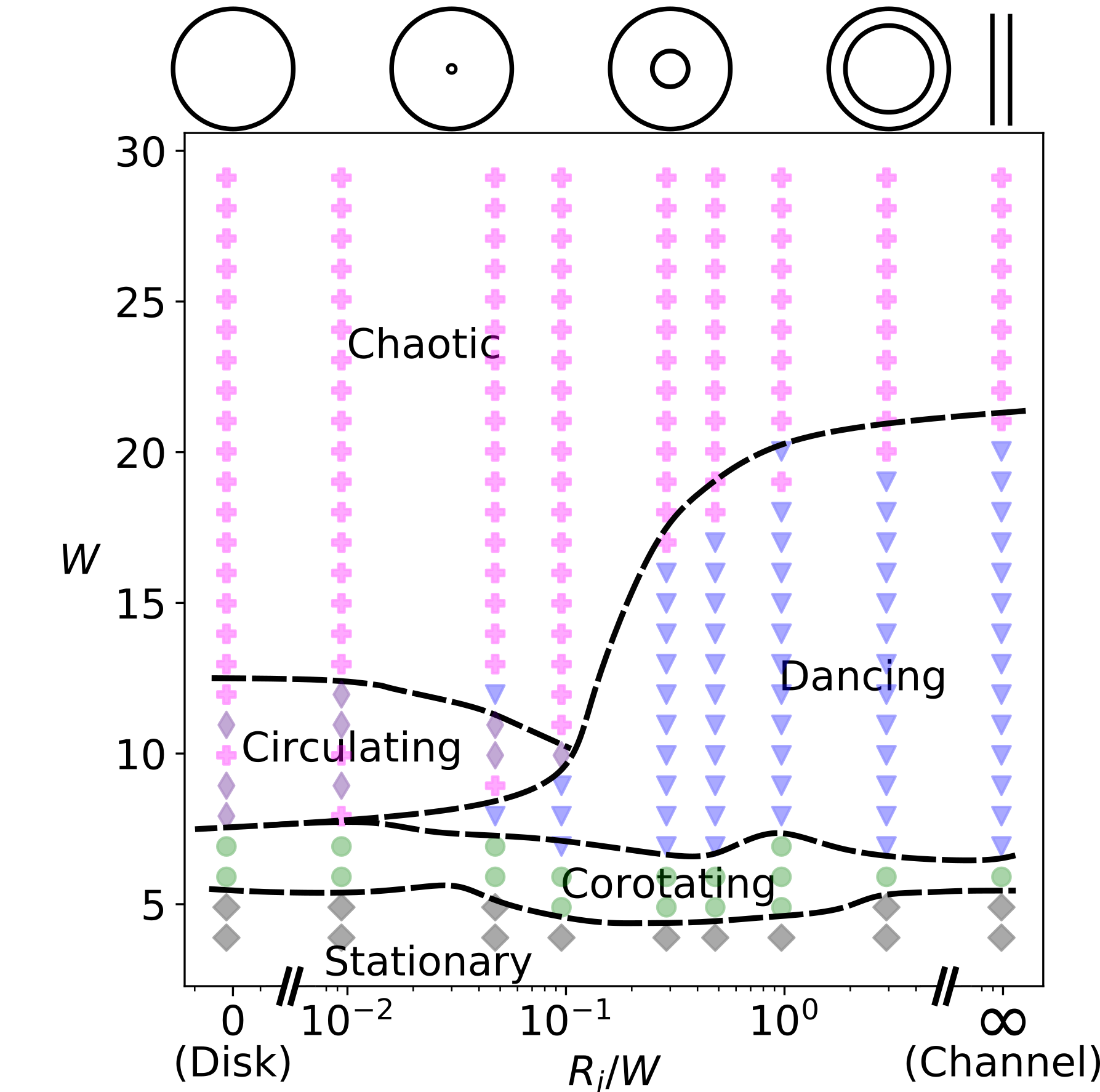}
\caption{
\textbf{Shape-size phase diagram for active nematics in annuli}. Steady states as a function of width $W$ and the ratio of the inner radius to width $R_i/W$: (i) Stationary state (grey diamonds), (ii) Corotating state (green circles), (iii) Circulating state (indigo thin diamonds), (iv) Dancing state (blue triangles), and (v) Chaotic state (magenta pluses). The indicated shapes for various $R_i/W$ on the top are a guide to the eye and not to scale. $R_i/W=0$ responds to a disk of radius $W$, whereas $R_i/W=\infty$ corresponds nominally to a channel simulation with width $W$, length $20W$, and periodic boundary conditions at the channel ends. For these simulations, we apply a no-slip boundary condition for the velocity, and weak parallel anchoring for the director, with $\EA=0.01$.
}
 \label{fig:phase_diagram}
\end{figure}

In this work we use hydrodynamic simulations to systematically investigate 2D active nematics confined to the annulus geometry. We vary the confinement shape by changing the inner radius of the annulus for a fixed width, going from a `disk-like' annulus, with a small pinhole at its center, to a `channel-like' annulus for which the radius is much larger than the channel width (Fig. ~\ref{fig:phase_diagram}). We vary the confinement size by changing the width of the annulus, and map out a shape-size phase diagram of the dynamical steady states. This phase diagram connects the previously observed steady states in disks and channels, and reveals how the system behaves between these two limits. In particular, the intermediate geometries reveal previously unobserved states exhibiting net material transport, which are stabilized by the positive curvature of the inner wall of the annulus. 

Further, we find that below a threshold inner radius, annuli exhibit the same dynamics as the corresponding true disks under a finite parallel anchoring on the nematic director, despite the two systems being constrained to have a different net topological charges by their boundary conditions. We provide a theoretical explanation for this observation and an estimate of the threshold pinhole size based on the competing anchoring and elastic energies at the hole. Finally, we find that the net-transport state disappears at high anchoring strengths. To understand these effects, we study the influence of activity on the boundary layer at the walls \cite{Norton2018,Hardouin2020a}, which uncovers a relationship between the anchoring strength and the curvature. More broadly, these results advance our understanding of how hydrodynamics and topological constraints of boundary conditions control the behavior of a confined active nematic.

\section{\label{sec:level2}Model Description}

As in our previous work \cite{Norton2018}, we study a continuum model of an active nematic described as a single incompressible fluid with internal nematic symmetry \cite{beris_edwards_1994,Doostmohammadi2018}. It has two fields:
the nematic tensor order parameter $\vb{Q} = \rho S [\vb{n}\otimes\vb{n}-(1/2)\vb{I}]$ and a flow
field $\vb{u}$. Here, $\vb{n}$ is the local orientation unit vector, $S$ is the scalar order parameter, and $\rho$ is the density, which we assume to be uniform and constant in the simulations throughout this paper. In the limits of low Reynolds number and high Ericksen number \cite{Koch2021}, the dynamics of this fluid is given by
\begin{equation}
    \eta \nabla^2 \vb{u} - \nabla P  - \alpha \nabla\cdot \vb{Q} = 0
\end{equation}
along with the incompressibility constraint $\nabla \cdot \vb{u}=0$. Here, $\eta$ is the dynamic viscosity and $P$ is the pressure. The third term is the leading order contribution from the extensile active stress $-\alpha\vb{Q}$, with $\alpha$ defined as the activity. We use no-slip boundary conditions on the velocity throughout this article, $\vb{u}|_{\partial\Omega}=0$. 

The dynamical equation for $\vb{Q}$ is given by
\begin{equation}
    \partial_t \vb{Q} + \nabla\cdot(\vb{u}\vb{Q}) = (\vb{Q}\Omega-\Omega\vb{Q}) 
    + \lambda \vb{E} + \gamma^{-1}\vb{H}
\end{equation}
Here, $\Omega_{ij} = (\partial_i u_j - \partial_j u_i)/2$ is the anti-symmetric vorticity tensor, $E_{ij} = (\partial_i u_j + \partial_j u_i)/2$ is the symmetric strain rate tensor, and $\lambda$ is the flow alignment parameter. The final term is the relaxation of the nematic field proportional to the variation of the nematic free energy, $H_{ij} = -\delta \mathcal{F} / \delta Q_{ij}$, with dissipation rate $\gamma^{-1}$. The nematic free energy for a confined system consists of a bulk contribution and a boundary contribution.

The bulk contribution is the Landau de Gennes free energy \cite{beris_edwards_1994} $\mathcal{F}_\text{LDG}$ given by
\begin{align*}
    \mathcal{F}_\text{LDG} = \int_{\Omega} \! &\dd^2\vb{r} 
    \left\{ C\qty(-\frac{\beta_1}{2} Q_{ij}Q_{ji} + \frac{\beta_2}{4}(Q_{ij}Q_{ji})^2)\right. \\
    &+ \left.\frac{1}{2}K(\partial_k Q_{ij}\partial_k Q_{ij}) \right\}.
\end{align*}
In this work, we take the simple form, $\beta_1(\rho)=\rho -1$ and $\beta_2(\rho)=(\rho+1)/\rho^2$ so as to set up the isotropic ($\rho<1$) to nematic ($\rho>1$) transition. We fix $\rho=1.6$ throughout this article, thus focusing on the far-from-critical nematic phase. At the boundary, we use the Nobili-Durand anchoring energy \cite{Nobili1992,Giomi2014}
\begin{equation*}
    \mathcal{F}_\text{ND} = \oint_{\partial \Omega} \! \dd \vb{r} \frac{1}{2} 
    \EA (Q_{ij}-W_{ij})(Q_{ji}-W_{ji}),
\end{equation*}
where $\vb{W}$ specifies the order and orientation at the boundary and $\EA'$ specifies the anchoring strength. Thus, the relaxation term becomes 
\begin{align*}
    \gamma^{-1}H_{ij} = &\Dr (\beta_1-\beta_2 Q_{kl}Q_{lk}) Q_{ij} + 2\DE \partial_k \partial_k Q_{ij} \\
    &- \EA (Q_{ij}-W_{ij})|_{\partial \Omega},
\end{align*}
where $\DE=(L_1+L_2)/2\gamma$, $\EA=\EA'/\gamma$ and $\Dr = C/\gamma$. 

We identify the time scale $T = 1/\Dr$ and the length scale $L=\sqrt{\DE/\Dr}$. We use these to non-dimensionalize the equations, with the dimensionless operators $\bar{\partial}_t = \partial_t / \Dr$ and $\bar{\partial}_i = \partial_i / \sqrt{\DE/\Dr}$. This gives the dimensionless equations
\begin{align*}
    \bar{\partial}_t \vb{Q} + &\bar{\nabla}\cdot(\bar{\vb{u}}\vb{Q}) = (\vb{Q}\bar{\Omega}-\bar{\Omega}\vb{Q}) 
    + \bar{\lambda} \vb{\bar{E}} + \bar{\vb{H}} \\
    \bar{H}_{ij} = &\Dr (\beta_1-\beta_2 Q_{kl}Q_{lk}) Q_{ij} + 2 \bar{\partial}_k \bar{\partial}_k Q_{ij} \\
    &- \bar{E}_A (Q_{ij}-W_{ij})|_{\partial \Omega}, \\
    &\bar{\nabla}^2 \bar{\vb{u}} - \bar{\nabla} P  - \bar{\alpha} \bar{\nabla}\cdot \vb{Q} = 0
\end{align*}
Here, the dimensionful quantities $\vb{u}$ and $P$ are re-scaled to $\bar{\vb{u}}=\vb{u}\sqrt{\DE \Dr}$ and $\bar{P}=P/(\Dr \eta)$. The dimensionful parameters get re-scaled as $\bar{E}_A=\EA/\Dr$, $\bar{\lambda}=\lambda/\Dr$, $\bar{\alpha}=\alpha / (\eta \Dr)$ and finally the system dimensions $\bar{R}=R/\sqrt{\DE/\Dr}$. In the following, we drop the overbars on the parameters and implicitly assume this nondimensionalization.

We start the simulations with zero velocity and uniform director field with 1\% noise. We use the open source Python library FEniCS \cite{AlnaesBlechta2015a, LoggWells2010a} to integrate the equations. The visualizations are made using the open source visualization software Paraview \cite{AhrensJames2005} or MATLAB \cite{MATLAB:2010}.
Unless specified otherwise, we use $\EA=0.01$ (weak anchoring) and $\alpha=\alpha_0=5$ (strong activity) in the simulations.
For this activity, the system reaches steady state within 500 simulation time units even for the largest geometries investigated. Accordingly, we simulate all the systems for a total of $\tf=1000$ time units and use the data from the last 10\% of the time for analysis. If we change the activity, we scale $\tf$ : $\tf' = (\alpha_0/\alpha)\tf$. For all simulations, we set $\dd t = 0.1$, and we use a mesh with average linear element size $\Delta x \sim 0.2$.

\begin{figure}[t]
    \centering
    \includegraphics[width=0.9\linewidth]{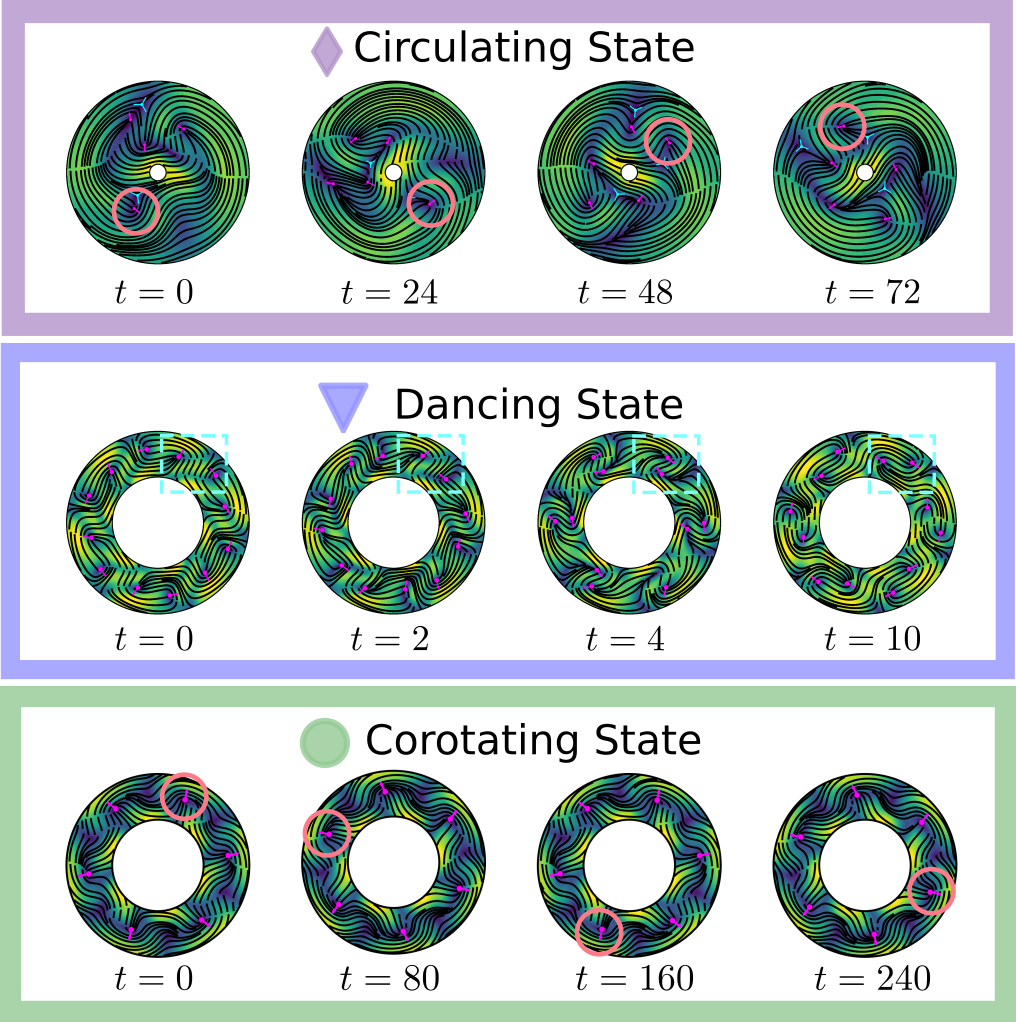}
    \caption{
    \textbf{Dynamical steady states in active nematics confined to annulus geometries.} (top) The circulating state, as observed for $W=10$ and $\Riw=0.1$. A single defect going around the channel is shown with a red circle to highlight the circulation. (middle) The dancing state, as observed for $W=8$ and $\Riw=1$. The highlighted region shows the dancing behavior of a defect pair over time. (bottom) The corotating state, as observed for $W=6$ and $\Riw=1$. The highlighted defect shows the corotating motion of the train of defects. The symbols correspond to those used to identify these states in the phase diagram in Fig. 3. All times start at an arbitrary point and are in the units of 0.1 simulation times. }
    \label{fig:illustrations}
\end{figure}

\section{\label{sec:level3}Results}
\subsection{Steady State Behaviors}
\label{sec:steady_state}

We specify the size of the annulus by its width $W$, and its shape with the ratio of the inner radius to the width, $\Riw$. We use $\Riw$ values from $0.01$ (disk-like) to $3$ (channel-like), and widths in the range $W \in [4,30]$ (see Fig. \ref{fig:illustrations} for some examples). For completeness, we also simulate active nematic disks and flat periodic channels. Here, the disk of radius $R$ is thought of as an ``annulus" with $W=R$ and $\Ri=0$, whereas a channel is thought of as an ``annulus'' with width $W$ and $\Ri\to\infty$. We thus get a shape-size phase diagram for the 2D active nematic confined to an annulus (see Fig. \ref{fig:phase_diagram}). We consider weak anchoring ($\EA=0.01$) and strong activity ($\alpha=5$). The system exhibits distinct dynamical steady-states that vary as a function of the  geometric parameters. Sequences of snapshots characterizing the dynamics for some of these states are shown in Fig. \ref{fig:illustrations}. Fig.~\ref{fig:phase_diagram} gives a phase diagram of the observed steady states. The specific criteria that we used to identify and classify these states are given in Appendix~\ref{app:classification}.

\paragraph{Stationary State} 
At very high confinements, active instabilities are suppressed, resulting in a class of stationary steady states, which we define as states for which the director field and velocity profile are independent of time: $\partial_t \vb{Q}=\partial_t \vb{u}=0$. The texture of the stationary steady state in the disk-like limit depends on the confinement geometry and boundary conditions, resulting in familiar dipolar states with two $\phd$ defects that closely resembles the behavior in a true disk \cite{Norton2018}, as well as spiral states with a time-independent non-zero velocity. We elaborate on the set of observed stationary states in Appendix~\ref{app:classification}. 
In the opposite high confinement extreme, \ie the periodic channel ($\Riw\rightarrow \infty$), we observe a uniform nematic with no flow. 

\paragraph{Corotating State}
Upon moving to weaker confinement by increasing $W$, the system transitions into a corotating state (see Fig.~\ref{fig:illustrations} (bottom)),
characterized by a train of isolated $\phd$ defects  circulating along the annulus. These defects always nucleate from the inner wall of the annuli. This is distinct from the straight periodic channel of the same width, where the bend-instability occurs from both boundaries resulting in a structure that is symmetric across the channel, although also with a spontaneously chosen circulation direction (see Movie S1).  The weak anchoring strength at the boundary allows for the corresponding $\mhd$ defects at the wall to exist. Thus, this state is natural for Neumann boundary conditions on $\vb{Q}$ with no anchoring. 
Fig.~\ref{fig:phase_diagram} shows that the corotating state occurs for a narrow range of $W$, with phase boundaries that are essentially independent of $\Ri$. This reflects the fact that the corotating state occurs when only a single $\phd$ defect can form across the narrowest confinement dimension (channel width in this case) \cite{Norton2018}. 

The simplest corotating state is observed at very small inner radii and consists of two $\phd$ defects rotating around the center and no $\mhd$ defects \cite{Norton2018,Opathalage2019}. Notably, this state is consistent with the net topological charge of $+1$ imposed by parallel anchoring in a true disk, but deviates from the net topological charge of 0 for an annulus. This indicates that, despite changing the topology, inserting a hole at the center of the disk does not affect the steady-state behavior below a threshold radius. As will be discussed in Section~\ref{subsec:topology}, the threshold size of the hole is expected to depend on the anchoring strength.
For larger inner radii, the $\mhd$ defects at the inner wall are stabilized by weak anchoring conditions and/or high inner curvature ($\Ri\rightarrow 0$). Thus, the corotating state is absent for low inner curvatures when the anchoring strength is high (discussed in Section~\ref{sec:anchoring}). 

\paragraph{Circulating state}
\begin{figure}
    \centering
    \includegraphics{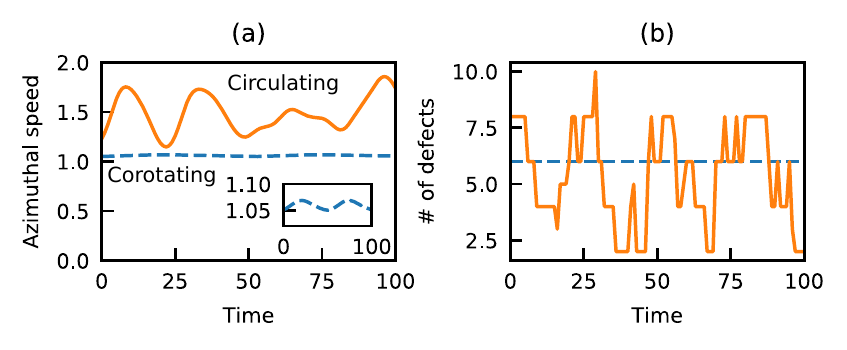}
    \caption{\textbf{Distinction between the corotating and the circulating states.} (a) Average azimuthal speed ($\vb{u}\cdot\hat{\theta}$) as a function of time for a representative corotating state (blue dashed line, $\Ri=6, W=6$) and a circulating state (orange continuous line, $\Ri=1, W=10$). While both have a steady circulation on average, there is significant variation in the instantaneous flow of the circulating state, as opposed to the weakly sinusoidal variation in the corotating state (the inset shows a zoomed-in plot of the same). (b) Total number of defects as a function of time for the same two datasets. The number of defects remains constant for the corotating case, whereas it fluctuates for the circulating case, indicating annihilation and proliferation.}
    \label{fig:circ_vs_cor}
\end{figure}
Upon increasing the width of the annulus in the disk-like limit ($\Riw \lesssim 0.1$), we observe a circulating state that is different from the corotating state, in that it exhibits turbulent-like nucleation and annihilation of defects in the steady state (see Fig.~\ref{fig:illustrations} (top)). However, the circulating and corotating states have similar vortex structures, with high circulation around a central vortex, albeit with significantly larger perturbations of the velocity profile in the circulating state (see Fig.~\ref{fig:circ_vs_cor}).

\paragraph{Dancing state}
For intermediate widths and relatively small curvatures, we observe the `dancing state', which has been well studied in the channel geometry \cite{Shendruk2017,Hardouin2019,Varghese2020} (see Fig.~\ref{fig:illustrations} (middle)). Here, pairs of $\pm 1/2$ defect pairs swim on a lattice of vortices along the channel. We find that the dancing state persists for annular channels with fairly high curvature (see Fig \ref{fig:phase_diagram}) as also noted in \cite{Shendruk2017}.

In the limit of low curvature ($\Riw\gtrsim 1$), the dancing state in the annulus closely resembles that observed in straight channels. In particular, there is usually an equal number of $\phd$ defects traveling in each direction along the channel axis, resulting in no net flow. This symmetry can be broken by formation of a `drift-lattice defect' \cite{Shendruk2017}, where two more $\phd$ defects travel in one direction than the other, resulting in a net drift. The probability of having such drift-lattice defects increases with the length of the channel \cite{Shendruk2017} or correspondingly the annulus circumference. 
\begin{figure}
    \centering
    \includegraphics{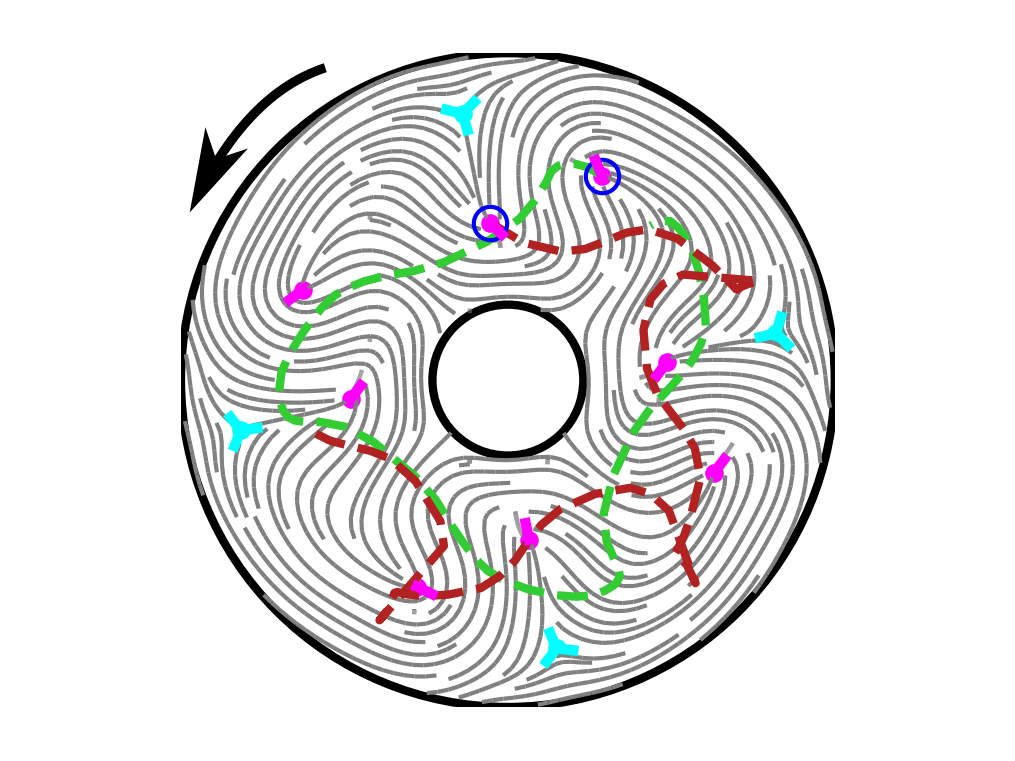}
    \caption{\textbf{Curvature-induced drift in the dancing state for $W=15$ and $\Riw=0.3$.} In this case, due to the asymmetry of the inner and outer curvature, the flow acquires a drift (see Movie S2) despite there being an equal number of defects moving in each direction. The dotted green line highlights the trajectory of a defect moving uninterrupted anti-clockwise, while the red dotted line highlights the trajectory of a defect moving clockwise, undergoing  annihilation and nucleation periodically at the outer edge. This asymmetry of defect motions creates a net flow in the counter-clockwise direction.}
    \label{fig:curvdrift}
\end{figure}
However, at high curvatures, $\Riw \lesssim 1$, the difference in the outer and the inner curvature results in an asymmetry between the $\mhd$ defects at those respective boundaries. This sometimes results in an asymmetry in the $\mhd$ defect interactions of the clockwise and anti-clockwise moving $\phd$ defects. The defects moving in one direction interact strongly with the outer $\mhd$ defects, which slows their flow, whereas the defects traveling in the other direction move freely. This results in a net `curvature-induced drift', even with the same number of defects moving in each direction (see Fig.~\ref{fig:curvdrift} and Movie S2). This behavior is unique to the annulus. 

\paragraph{Chaotic state}
Finally, above a threshold width, whose value depends on the curvature, the system undergoes extensive defect proliferation resulting in turbulent -like dynamics and no coherent motion or circulation. 

\subsection{Effect of topology}
\label{subsec:topology}

The similarity between the steady-states in the disk and pin-hole annuli discussed for the corotating state extends for all widths (Fig.~\ref{fig:phase_diagram}). That is,  the steady state dynamics are insensitive to the presence of a pin-hole below a threshold value of the inner radius. 
The origin of this insensitivity can be understood from the following. First, near the pin-hole, the energy cost for a uniform director (resulting in the insensitivity) goes as $E_\text{uniform} \sim \EA \Ri$, whereas the energy for a perfectly anchored ($+1$ defect) structure goes as $E_\text{defect} \sim K \log (\rho_\text{max}/\Ri)$, where $\rho_\text{max}$ is a characteristic defect spacing or confinement size\cite{degennes_prost_1993}. The pin-hole will only affect the topology of the  director field if $E_\text{uniform} > E_\text{defect}$. Thus, for sufficiently small holes, the director structure will be indistinguishable from that of a  disk. Second,
in the annulus, the no-slip boundary condition on the fluid at the boundary of the hole imposes zero fluid velocity. While this requirement is absent in the disk, the symmetry of the corotating state results in a vanishing velocity at its center. Thus, the vortex structures of the disk and annulus with small $\Ri$ are very similar.
These two effects combine to make the system dynamics topologically insensitive to the pinhole.

\subsection{\label{sec:anchoring}Effect of anchoring strength}
\begin{figure}
\centering 
\includegraphics{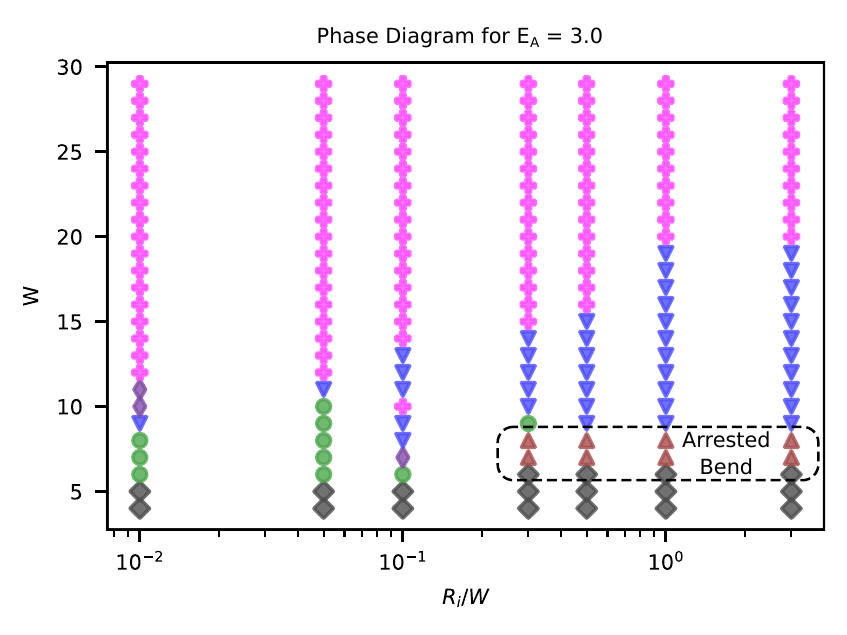}
\caption{\textbf{Phase diagram for strong anchoring conditions} (with $\EA=3.0$). The markers correspond to the same states as in Fig. \ref{fig:phase_diagram}: (i) Stationary state (grey diamonds), (ii) Corotating state (green circles), (iii) Circulating state (indigo thin diamonds), (iv) Dancing state (blue triangles), and (v) Chaotic state (magenta pluses), except for the maroon upward triangles that indicate the arrested bend instability state. This phase diagram reveals the absence of the corotating state (green circles) for high anchoring and low curvature.}
\label{fig:phase_diagram_3}
\end{figure}

\begin{figure}
    \centering
    \includegraphics[width=\linewidth]{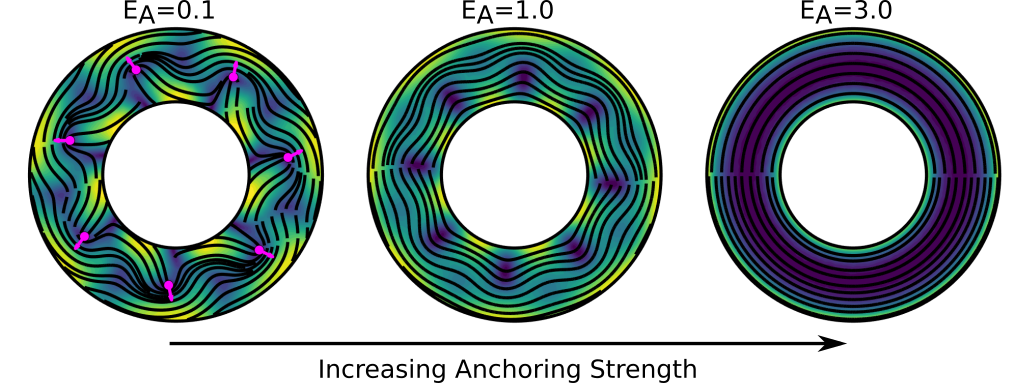}
    \caption{\textbf{The corotating state disappears at high anchoring strengths.} Snapshots of the steady state for $W=6$, $\Riw=1$ for different anchoring strengths. (left) For low anchoring strengths, the $\mhd$ defects are easier to form and we get the corotating state. (middle) For intermediate anchoring, the bend instability is arrested. (right) For high anchoring strengths, the bend instability is completely suppressed. The width for the onset of the dancing state remains independent of anchoring strength.}
    \label{fig:corotating_anchoring}
\end{figure}

The results described thus far have focused on a relatively weak anchoring strength $\EA=0.01$. Fig. \ref{fig:phase_diagram_3} shows a corresponding phase diagram for $\EA=3$. The key differentiating feature of this phase diagram is that the corotating state disappears in the straight-channel limit. In this regime, as $W$ increases, the system transforms from a uniform nematic with azimuthal orientation, to an arrested bend instability, and then to the dancing state. This can also be seen by increasing the anchoring strength for a fixed $W$ and $\Ri$ (Fig.~\ref{fig:corotating_anchoring}). We note this observation is similar to results from simulations of polar active fluids confined to channels, where coherent transport through the channel is suppressed at high anchoring strengths \cite{Gulati2022}. This suggests that it should be explainable by a simple argument. To this end, we note that the altered phase diagram in our system reflects that the formation of the corotating state in the channel limit requires two conditions: a width that accommodates only one defect ($W\leq \ell_\text{d}$, section~\ref{sec:steady_state})  and  a sufficiently small energy cost to form the corresponding $\mhd$ defects at the inner wall. Above a threshold value of $\EA$, this energy cost is too large for channel-like annuli, and thus the bend instability is arrested before defect nucleation can occur. For smaller inner radii, $\mhd$ defects are stabilized by the curvature of the inner wall and thus incur a smaller energy. Moreover, in the disk-like limit $\Riw \lesssim 0.1$, the corotating state forms without any $\mhd$ defects. Hence, the corotating state emerges at smaller inner radii.

\begin{figure*}[t]
\centering
\includegraphics[width=0.9\linewidth]{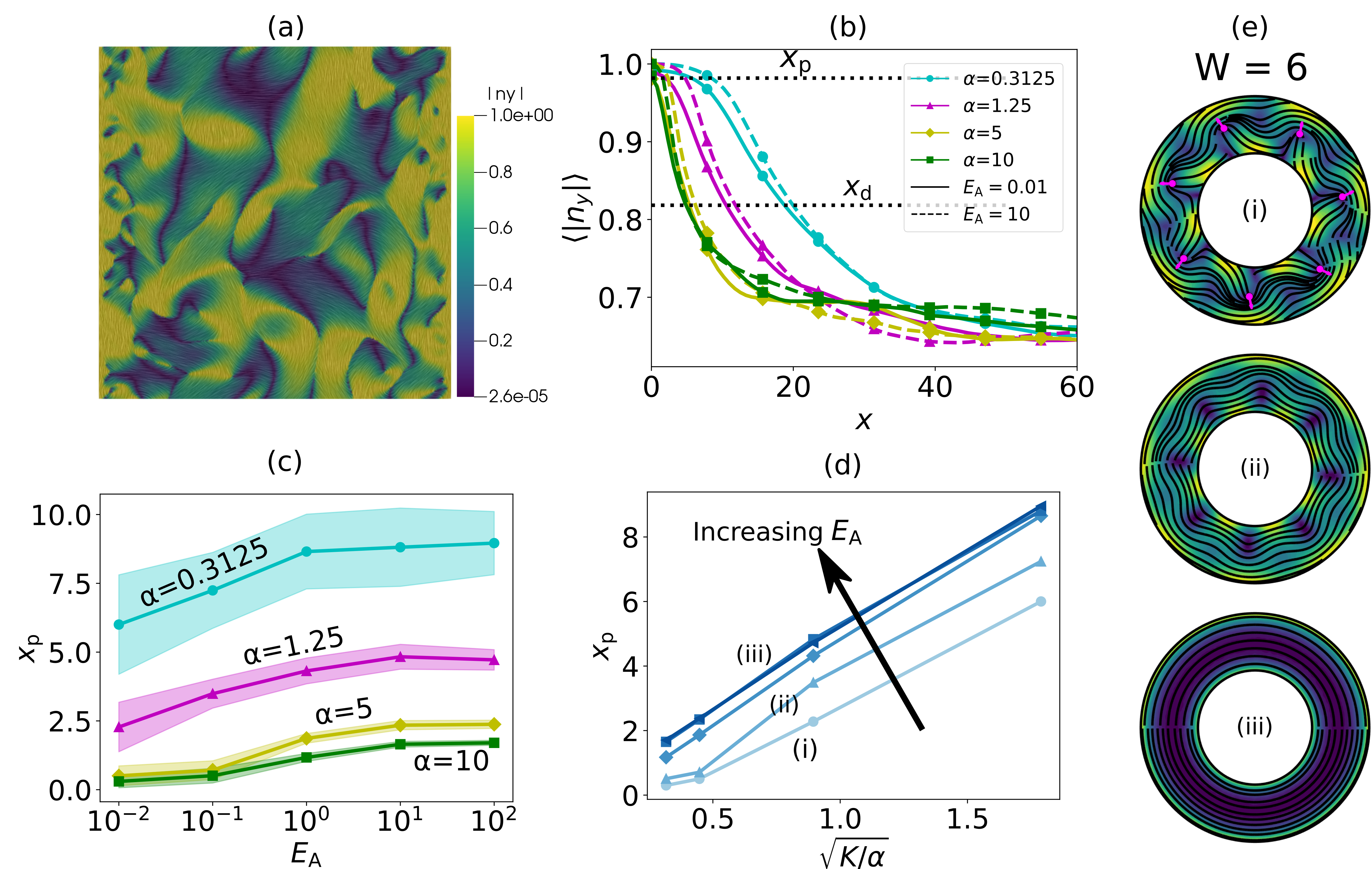}
\caption{\textbf{The effects of anchoring are renormalized by activity}. (a) Snapshot of a simulation with $\alpha=5$ and $\EA=3$. The $200\times 200$ box has periodic boundary conditions in the $y$ direction and parallel anchoring with strength $\EA$ in the x direction. The color shows $|n_y|$. (b) $\expval{|n_y|}$ as a function of $x$ for the steady state, averaged over time and $y$ for various activities, each for low (solid lines) and high (dashed lines) anchoring cases. The dotted lines indicate the thresholds for the definition of the alignment plateau length $\xp$ and decay length $\xdecay$. (c) $\xp$ vs. $\EA$ for four different activities. The shaded region indicates the standard error. (d) $\xp$ as a function of the active length scale $\sqrt{K/\alpha}$, for various anchoring energies. The data in this plot is same as that in (c), with the anchoring energies increasing with the darkness of the lines. Points marked (i), (ii) and (iii) highlight specific parameter sets. (e) Steady states obtained using the parameters in (d) (i), (ii), and (iii).}
\label{fig:anchoring}
\end{figure*}
To understand the effect of the anchoring strength on the phase behavior and in particular on the existence of the corotating state, we investigated how the penetration length of the parallel anchoring boundary condition depends on control parameters in an active nematic. To this end, we performed simulations in a square channel geometry, with dimensions $200\times200$, periodic boundary conditions in the $y$ direction and parallel anchoring boundary conditions in $x$ (See Fig.~\ref{fig:anchoring}). Note that this domain is much larger than the systems investigated in Fig.~\ref{fig:phase_diagram}, and the active nematic exhibits turbulent-like behavior everywhere except near the vertical boundaries. 

To quantify the effect of anchoring in this system, Fig.~\ref{fig:anchoring} shows the extent to which the alignment with the vertical wall persists into the channel interior. In particular, we have plotted $\nyexp$ as a function of $x$, where $|n_y|$ is the magnitude of the vertical component of the director field, and the average is performed over the vertical dimension $y$ and time $t$. Strong parallel anchoring implies $|n_y(x=0)| \approx 1$, whereas in the bulk we expect a uniform distribution of $n_y$, yielding $\nyexp(x\gg1) = \expval{|\sin\theta|} = 2/\pi \approx 0.64$. For convenience, we define a parameter indicating the fractional drop in $\nyexp$ from 1 towards the bulk value $2/\pi$ as $\Deltany(x) = \frac{1-\nyexp(x)}{1-2/\pi}$.

As shown in Fig.~\ref{fig:anchoring}, we observe that the variation of $\nyexp$ away from the vertical wall exhibits two characteristic length scales. First, there is a plateau within which the director is parallel to the wall $\nyexp\approx 1$, followed by an exponential decay to its bulk value. We define the width of the plateau region, $\xp$, as the distance from the wall at which alignment diminishes by 5\%, $\Deltany(\xp)=0.05$, and the decay length $\xdecay$ as the distance corresponding to 50\% of the alignment decrease, $\Deltany(\xdecay)=0.5$ (see Fig.\ref{fig:anchoring}(b)).

\begin{figure}
    \centering
    \includegraphics[width=\linewidth]{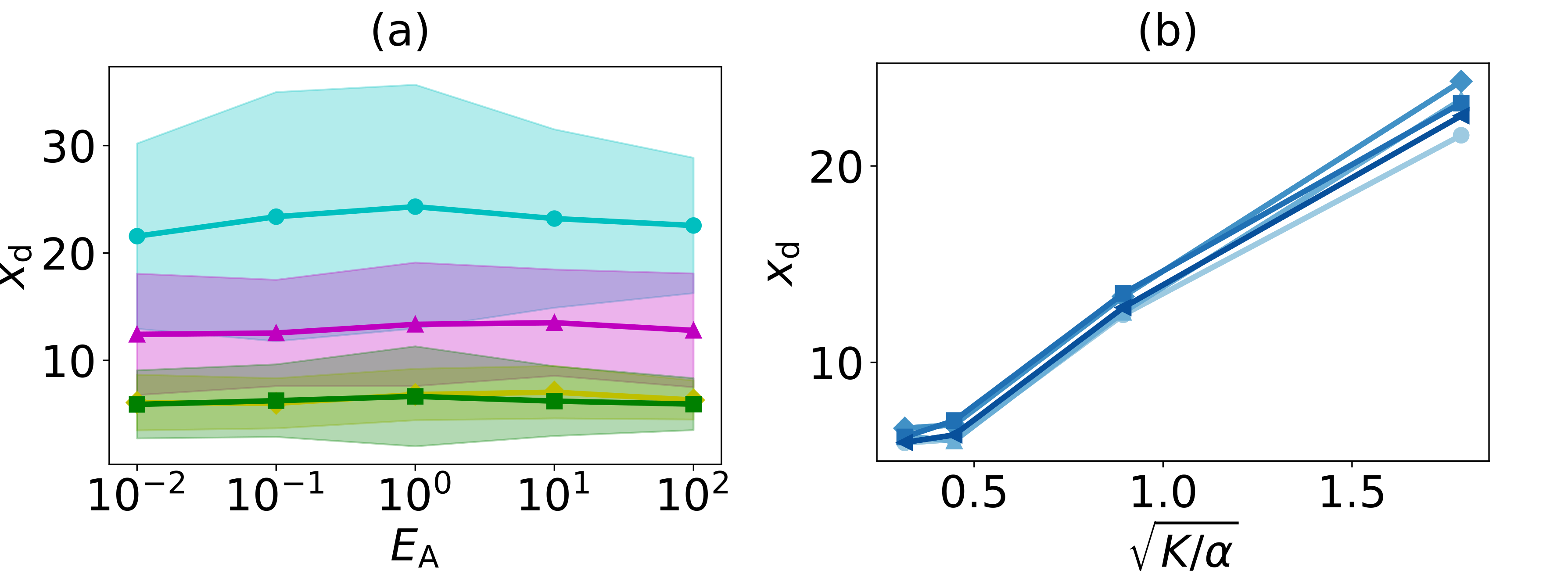}
    \caption{\textbf{Alignment decay length $\xdecay$ as a function of activity and anchoring strength.} Plot of $\xdecay$ (defined in Fig.~\ref{fig:anchoring}) as a function of anchoring strength and the active length scale, analogous to Fig.~\ref{fig:anchoring}(c) and (d), indicating the $\xdecay$ is independent of the anchoring strength.}
    \label{fig:xd_vs_EA}
\end{figure}

We find that the overall decay length is primarily determined by activity. In particular, both the plateau width $\xp$ and decay length $\xdecay$ increase linearly with the active length scale $\sqrt{K/\alpha}$ \cite{Giomi2015} and thus diminish with increasing activity. In contrast, $\xdecay$ is essentially independent of anchoring strength, while $\xp$ increases moderately with $\EA$ until saturation (Fig.~\ref{fig:anchoring}(c,d), Fig.~\ref{fig:xd_vs_EA}). We now compare the value of $\xp$ with the channel width for the corotating state, $W=6$ (see Fig.~\ref{fig:anchoring}(e) and points (i-iii) on Fig.~\ref{fig:anchoring}(d)). For the corotating state at $W=6$ and $\EA=0.01$, $\xp\approx0.5$, whereas for $\EA=1$, $\xp\approx 1.86$. For the higher anchoring strengths, the plateau region covers roughly $60\%$ of the entire annulus, thus hindering defect formation, whereas for the lower anchoring case, it covers only $\sim 15\%$ of the annulus, thus allowing for defect nucleation. Importantly, the anchoring boundary layer only affects the director profile and not the velocity profile. A similar (no-slip) layer in the velocity profile would have merely shifted the width value at which the corotating state occurs, as opposed to eliminating it. These observations thus directly explain the findings in Fig.~\ref{fig:corotating_anchoring}. More broadly, the small value of $\xp$ for high activities explains the observations from this work and Norton \etal \cite{Norton2018} that strongly confined active nematics are insensitive to topological constraints, and that the anchoring effects are restricted to a narrow region near the boundary.

\section{\label{sec:discussion}Discussion}
Using an annulus confinement, we simulate the natural periodic boundary conditions one can build in experiments, while studying the role of curvature in the steady state dynamics. We map a shape-size phase diagram for annular active nematics. At high confinement, we observe  a set of stationary states (meaning that the director field and velocity profiles are independent of time, see Appendix~\ref{app:classification}). As we progressively reduce confinement or increase activity, we find four distinct dynamic steady states as follows. First, we observe a corotating (coherently-flowing) state with a train of defects that are stabilized by the $\mhd$ defects staying near the inner curved wall. This state is unique to the annulus geometry. Second, we observe a circulating state, which exhibits a vortex structure similar to a corotating disk state but with significant perturbations. Third, we observe the dancing state that is well known to occur in channel-like geometries \cite{Shendruk2017}. However, within this regime we observe a behavior that is unique to the annulus geometry, in which the curvature of the inner wall drives and asymmetry of defect motions in opposite directions, leading to a net `curvature-induced drift'. Finally, at low confinement or high activity we observe a chaotic state.  

We observe that a sufficiently small hole in the disk, while changing the topology, does not change the dynamical steady state of the active nematic. We present scaling analysis that suggests that this threshold results from a competition between the anchoring energy and elastic energy of the nematic. Consistent with this analysis, the threshold pinhole size below which the system dynamics is insensitive to topology decreases with increasing anchoring strength. Thus, the behavior of a pinhole annulus can be globally switched by tuning the boundary anchoring or its geometry (pinhole size).
Lastly, we also explore the effect of boundary anchoring on the steady states, finding that a large anchoring strength destabilizes the $\mhd$ defects in the corotating state. Probing the influence of the anchoring on the boundary layer in bulk systems, we find that activity depletes the boundary layer. This finding explains a previous observation that the director fields of active nematics confined in disk geometries are insensitive to boundary conditions \cite{Norton2018}, and furthermore extends that analysis to additional geometries. Similar analyses can be applied to any geometry in which topological constraints imposed by the boundaries complete with the preferred global arrangement of the director field.

We note that our model does not capture all the states that are observed in experiments of microtubule-kinesin active nematics in annulus geometries \cite{Zarei2021}. Additional features of the experimental system that could be added to the model include the following.  In the experiments the microtubule nematic floats at an oil-water interface and is thus coupled to two isotropic fluids, while the theory approximates the system as a single component fluid. While the no-slip velocity boundary conditions of the theory are consistent with the oil/water fluid, the microtubules can slip at the boundaries. Further, the microtubule bundles have finite thickness and length, and thus can act as material lines that exert non-local forces \cite{Opathalage2019}. In the continuum model, the director field only exerts stress locally. Extending the model to incorporate these effects could address the discrepancies between theory and experiment.

\section*{Conflicts of interest}
There are no conflicts to declare.

\begin{acknowledgements}
This work was supported by the National Science Foundation (NSF) DMR-1855914 (CJ and MFH), DMR-2202353 (AB) and the Brandeis Center for Bioinspired Soft Materials, an NSF MRSEC DMR-2011846 (CJ, AB, and MFH). Computing resources were provided by the NSF XSEDE/ACCESS allocation MCB090163 (Stampede and Expanse) and the Brandeis HPCC which is partially supported by NSF DMR-MRSEC 2011846 and OAC-1920147. 
\end{acknowledgements}

\appendix
\section{Classification of steady-states}
\label{app:classification}
To aid the classification of the steady-states, we define a signed order parameter indicating the degree of flow, $\Phi(t) = \expval{\vb{u}\cdot\hat{\vb{e}}_{\theta}/|\vb{u}|}$ following Opathalage \etal \cite{Opathalage2019}. Using this order parameter and other aspects of the dynamics, we categorize the steady states as follows:
\paragraph{Stationary state}
\begin{figure}
    \centering
    \includegraphics{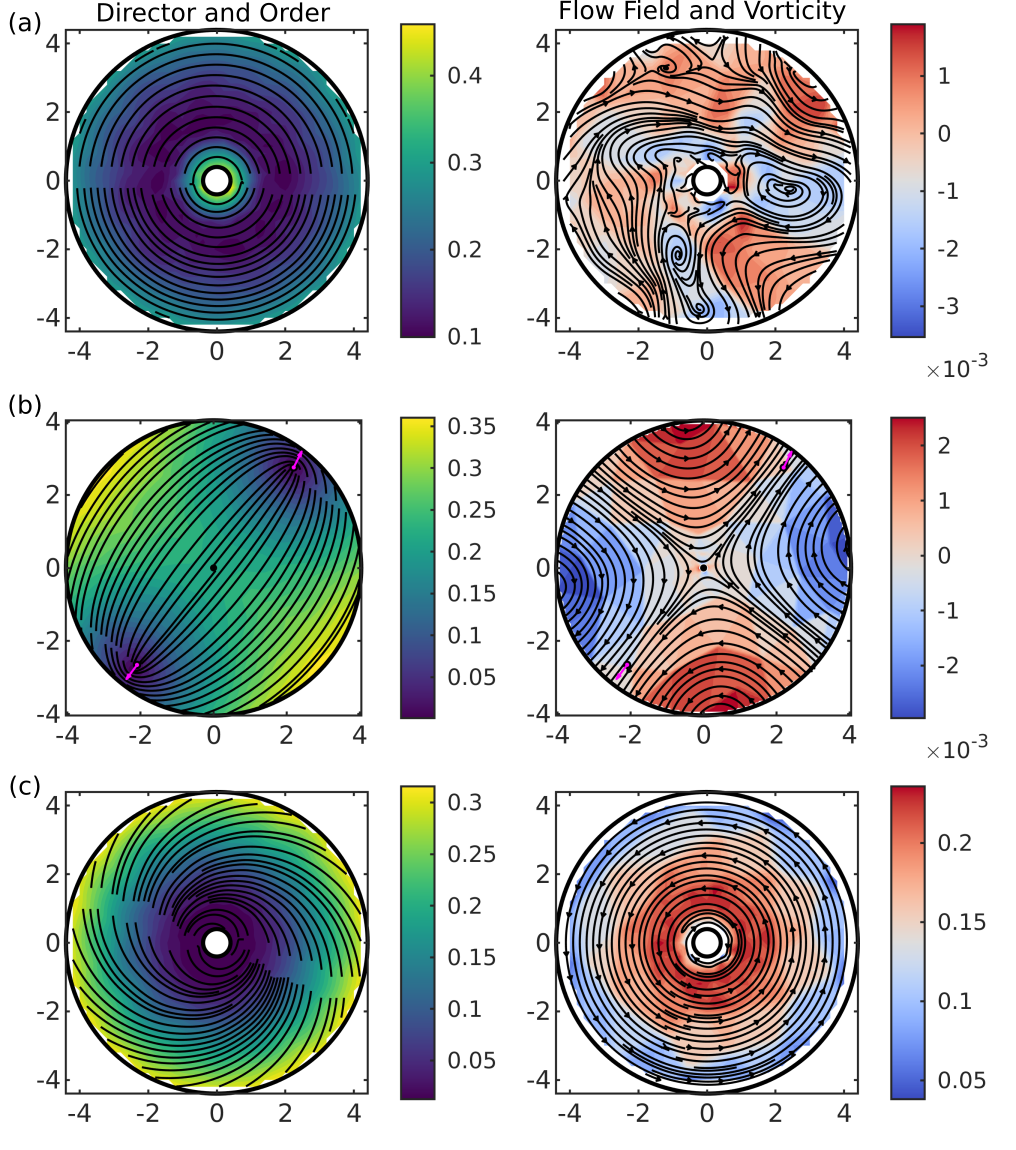}
    \caption{\textbf{Various stationary states ($\partial_t \vb{Q}=0$) of active nematics observed in disk-like annuli.} The left panel shows the director field with streamlines and the scalar order parameter $S$ with the color. The right panel shows the corresponding flow field, also shown as directed streamlines and the vorticity $|\nabla \times \vb{v}|$ with the color. (a) State with $\vb{v}\sim 0$ ($|\nabla\times \vb{v}|\sim 10^{-3}$) and the director in a $+1$ defect configuration. Here, $\EA=3, W=4, \Riw=0.1$. (b) Dipolar state with two defects and $v\sim 0$ ($|\nabla\times v|\sim 10^{-3}$), as also observed in disks in Ref.~\cite{Norton2018}. Here, $\EA=1, W=4, \Riw=0.01$. (c) For low anchoring strength, we can get a spiral configuration of the nematic, where we get a non-zero but constant velocity ($
    \vb{v} \neq 0, \partial_t \vb{v}=0$). Here, $\EA=0.01, W=4, \Riw=0.1$.}
    \label{fig:stationary}
\end{figure}
A state with a time-independent director field and velocity profile ($\partial_t \vb{Q}=0=\partial_t \vb{u}$) is labeled as a stationary state. Depending on the confinement and anchoring conditions, we observe different textures of the nematic, which in turn result in different velocity profiles (Fig.~\ref{fig:stationary}). High confinement and high anchoring strength result in a circular nematic profile corresponding to the $+1$ defect anchoring condition, with $\sim 0$ velocity, as seen in Fig.~\ref{fig:stationary}(a). For slightly lower confinement, we observe the dipolar state with two $\phd$ defects, as also observed in disks in Ref.~\cite{Norton2018} (Fig.~\ref{fig:stationary}(b)). An interesting state occurs at low anchoring, where the director aquires a spiral configuration, resulting in a stationary but \textit{non-zero} velocity, as shown in Fig.~\ref{fig:stationary}(c). This state is also peculiar because it has not been observed in the experiments to  our knowledge. 
\paragraph{Corotating state}
A state with defects revolving around the center of the annulus in a steady circulation is labeled a corotating state. Another defining property of the corotating state is that its $\vb{Q}$ and $\vb{u}$ profiles are constant in a rotating (or translating in the case of straight channels) frame of reference with the appropriate angular (or horizontal) speed.
\paragraph{Circulating state}
A state with net circulation ($\Phi(t)$ remaining far from 0 and not changing sign, see Fig.~\ref{fig:circ_vs_cor}) but with steady nucleation and annihilation of defects is labeled a circulating state. This is unlike the corotating state where the defects, once formed, remain in motion without annihilating.
\paragraph{Dancing state}
A state with pairs of defects swimming along vortices, as extensively characterized in Shendruk \etal \cite{Shendruk2017}.
\paragraph{Chaotic state}
A state exhibiting proliferation (steady nucleation and annihilation) of defects, with no net flow ($\expval{\Phi(t)}=0$) is labeled a chaotic state.

\paragraph{Arrested bend instability}
A state with the nematic undergoing bend instability, but unable to flow or exhibit the dancing pattern due to high confinement.
\section{Defect detection and tracking}

To locate the defects, we compute a map of the signed winding number $w = 1/(2\pi) \oint \nabla\theta \cdot \dd{\vec{s}}$ at every point in space \cite{Kamien2002,Norton2018} with an integration ring of radius of 5 pixels. The winding number is zero everywhere except at the defect locations \cite{DeCamp2015,Ellis2018}. To eliminate spurious defects, we filter out regions with a non-zero winding number that are smaller than 60 squared pixels in area. 

To plot the defect trajectories in Fig.~\ref{fig:curvdrift}, the $\phd$ defects are tracked using the open source software Trackpy \cite{Allan2021} using a \texttt{search\_range} value of 20 pixels. The trajectories thus obtained are further filtered with a threshold of minimum three frames of survival. 

%

\end{document}